\documentclass[11pt,preprint]{aastex}

\shorttitle{Multi-Thread Hydrodynamic Modeling}
\shortauthors{Warren}

\begin{document}


\title{Multi-Thread Hydrodynamic Modeling of a Solar Flare}

\author{Harry P. Warren}

\affil{E. O. Hulburt Center for Space Research, Code 7670, Naval
       Research Laboratory, Washington, DC 20375 \\
       hwarren@nrl.navy.mil}


\begin{abstract}
 Past hydrodynamic simulations have been able to reproduce the high
 temperatures and densities characteristic of solar flares. These
 simulations, however, have not been able to account for the slow
 decay of the observed flare emission or the absence of blueshifts in
 high spectral resolution line profiles. Recent work has suggested
 that modeling a flare as an sequence of independently heated threads
 instead of as a single loop may resolve the discrepancies between the
 simulations and observations. In this paper we present a method for
 computing multi-thread, time-dependent hydrodynamic simulations of
 solar flares and apply it to observations of the Masuda flare of 1992
 January 13. We show that it is possible to reproduce the temporal
 evolution of high temperature thermal flare plasma observed with the
 instruments on the \textit{GOES} and \textit{Yohkoh} satellites.  The
 results from these simulations suggest that the heating time-scale
 for a individual thread is on the order of 200\,s. Significantly
 shorter heating time scales (20\,s) lead to very high temperatures
 and are inconsistent with the emission observed by
 \textit{Yohkoh}. 
\end{abstract}



\section{Introduction}

 Solar flares are a potentially rich source of information on how
 energy is released during magnetic reconnection. At present, however,
 our physical understanding of solar flares is largely
 qualitative. Solar flare models based on magnetic reconnection are
 broadly consistent with observations, but detailed comparisons
 between numerical models and observations have been generally
 unsuccessful.  For example, most previous attempts to model solar
 flares with time dependent hydrodynamic simulations have not been
 able to account for essential aspects of the observations, such as
 the evolution of the observed emission or the details of the spectral
 line profiles (e.g., \citealt{peres1987,mariska1991}). Hydrodynamic
 simulations indicate that high density flare plasma cools rapidly,
 while soft X-ray emission from solar flares often persists for many
 hours. Hydrodynamic simulations also predict that high velocity
 upflows should be observed during the impulsive phase of the
 flare. In the vast majority of observed line line profiles, however,
 the stationary component is dominant (e.g., \citealt{mariska1993})

 One difficulty with most previous hydrodynamic modeling efforts is
 that they have treated the flare as a single loop.  This is clearly
 inconsistent with observations taken at high spatial resolution
 which show that solar flares are not the result of depositing energy
 into a single loop. Rather, these observations suggest that solar
 flares are composed of many small-scale threads (e.g.,
 \citealt{warren1999b,warren2000b,aschwanden2001c}). Furthermore,
 flare observations also show that plasma at many different
 temperatures is present simultaneously indicating that these threads
 are not all heated at once, but over the duration of the event.

 Some recent work has attempted to incorporate this observational
 understanding of magnetic reconnection into hydrodynamic simulations
 and model flares as a superposition of many independently heated
 threads.  \cite{hori1997,hori1998}, for example, found that the
 spatially averaged line profiles resulting from a succession of
 independently heated loops will generally be dominated by the
 stationary component, in qualitative agreement with observations.
 However, they were not able to perform detailed comparisons with
 observations.  Using a similar approach \cite{reeves2002a} were able
 to model the temporal evolution of emission at both high temperatures
 ($\sim10$\,MK) and relatively cool temperatures ($\sim1$\,MK) during
 the rise phase of an event. Some discrepancies between the simulation
 and the observations were evident, however, during the decay of the
 event they modeled.  This work also did not use full solutions to the
 hydrodynamic loop equations, rather they employed the
 \cite{cargill1995} scaling laws to approximate the evolution of the
 plasma.

 Treating a flare as a succession of threads introduces additional
 complexities into the hydrodynamic modeling. Since the observed
 emission results from the superposition of many threads it is not
 obvious how to determine the properties of an individual
 thread. Recently, \cite{warren2005a} have shown how to use
 \textit{GOES} soft X-ray fluxes to derive the energy flux and volume
 for each thread using a minimum of assumptions. This algorithm allows
 for detailed comparisons between a multi-thread, time-dependent
 hydrodynamic simulation and observations. For the \ion{Ca}{19} and
 \ion{S}{15} resonance lines they found generally good agreement for
 both the intensity and the line profiles observed during the initial
 phase of a flare.  They found that the strongly blue-shifted emission
 evident during the initial heating of a thread is largely masked by
 threads that have been heated previously and do not show bulk
 motions. The extended lifetime of the flare relative to a
 characteristic cooling time was easily reproduced using a secession
 of threads.

 In this paper we present a detailed description of the algorithm for
 computing a multi-thread, time-dependent hydrodynamic simulation of a
 solar flare. We apply this modeling to a well studied event, the
 Masuda flare of 1992 January 13 (e.g., \citealt{masuda1995}).  Our
 focus here is on the evolution of the thermal emission observed with
 the Soft X-Ray Telescope (SXT), the Bragg Crystal Spectrometer (BCS),
 and Hard X-Ray Telescope (HXT) instruments on \textit{Yohkoh}.  Of
 particular interest is the evolution of the BCS \ion{Fe}{25} and HXT
 light curves. Emission observed in these channels is formed at very
 high temperatures and is likely to be the most sensitive to the
 details of the energy deposition during the flare.

\section{Observations}

\begin{figure*}[t!]
 \centerline{%
 \includegraphics[scale=0.585]{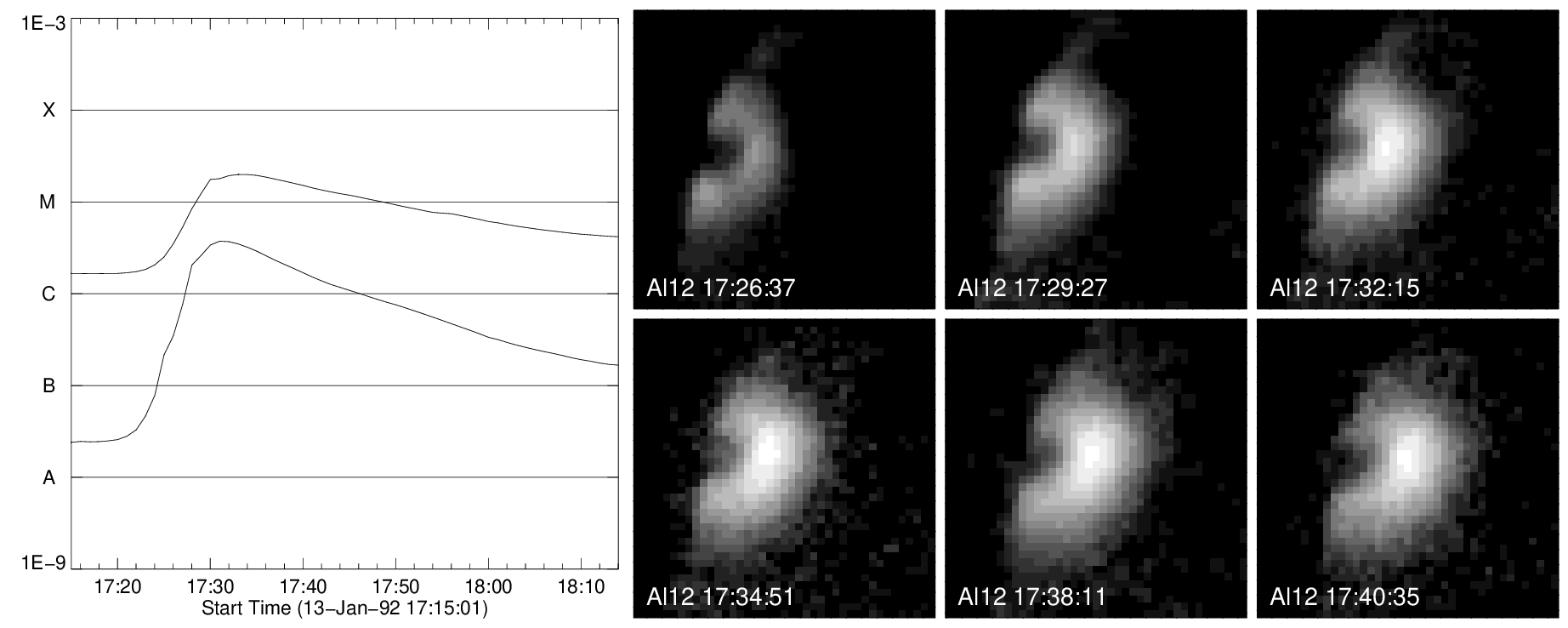}}
 \caption{Observations of the 1992 January 13 Masuda
 flare. (\textit{left panel}) \textit{GOES} fluxes in the 1--8 and
 0.5--4\,\AA\ channels. (\textit{right panels}) SXT images of the
 flare arcade in the thick aluminum filter. These images span the
 available SXT observations. Each image is scaled logarithmically
 using a common intensity scaling.}
 \label{fig:sxt}
 \end{figure*}

 The Masuda flare was a \textit{GOES} M2.0 flare that occurred on the
 west limb of the Sun on 1992 January 13. As is shown in
 Figure~\ref{fig:sxt}, the event began at about 17:22, peaked at about
 17:33, and decayed slowly over the next two hours. This flare was
 well observed by the instruments on the \textit{Yohkoh}
 spacecraft. The \textit{Yohkoh} instruments include the Soft X-ray
 Telescope (SXT, \citealt{tsuneta1991}), the Bragg Crystal
 Spectrometer (BCS, \citealt{culhane1991}), and the hard X-ray
 telescope (HXT, \citealt{kosugi1991}).

 The soft X-ray images from SXT for this event are shown in
 Figure~\ref{fig:sxt}. These images suggest a simple rising,
 two-ribbon flare arcade structure. The height of the arcade moves up
 by about 7\arcsec\ during the time the SXT images are available (from
 approximately 17:26 to 17:42). \cite{aschwanden1996} estimate the
 radius for the soft X-ray emission to be 12,500\,km at 17:28, which
 corresponds to a total loop length of about 39\,Mm.

 The emission in the lowest energy HXT channel (L 14--23 keV) is
 generally co-spatial with the loops imaged with SXT, indicating that
 the emission in this energy range is predominately thermal in origin.
 At higher energies (M1 23-33 keV and M2 33--53 keV) the emission
 emanates from the loop footpoints and a relatively weak source above
 the arcade (e.g., \citealt{masuda1995}). Additional hard X-ray
 observations of the Masuda flare are available from the BATSE
 instrument on the \textit{Compton Gamma Ray Observatory}
 (\textit{CGRO}). From these data \cite{aschwanden1996} compute a
 spectral index of $\gamma=3.74$ for the 30--120\,keV energy range
 near the peak of the hard X-ray emission.

\begin{figure*}[t!]
 \centerline{%
 \includegraphics[scale=0.585,angle=90]{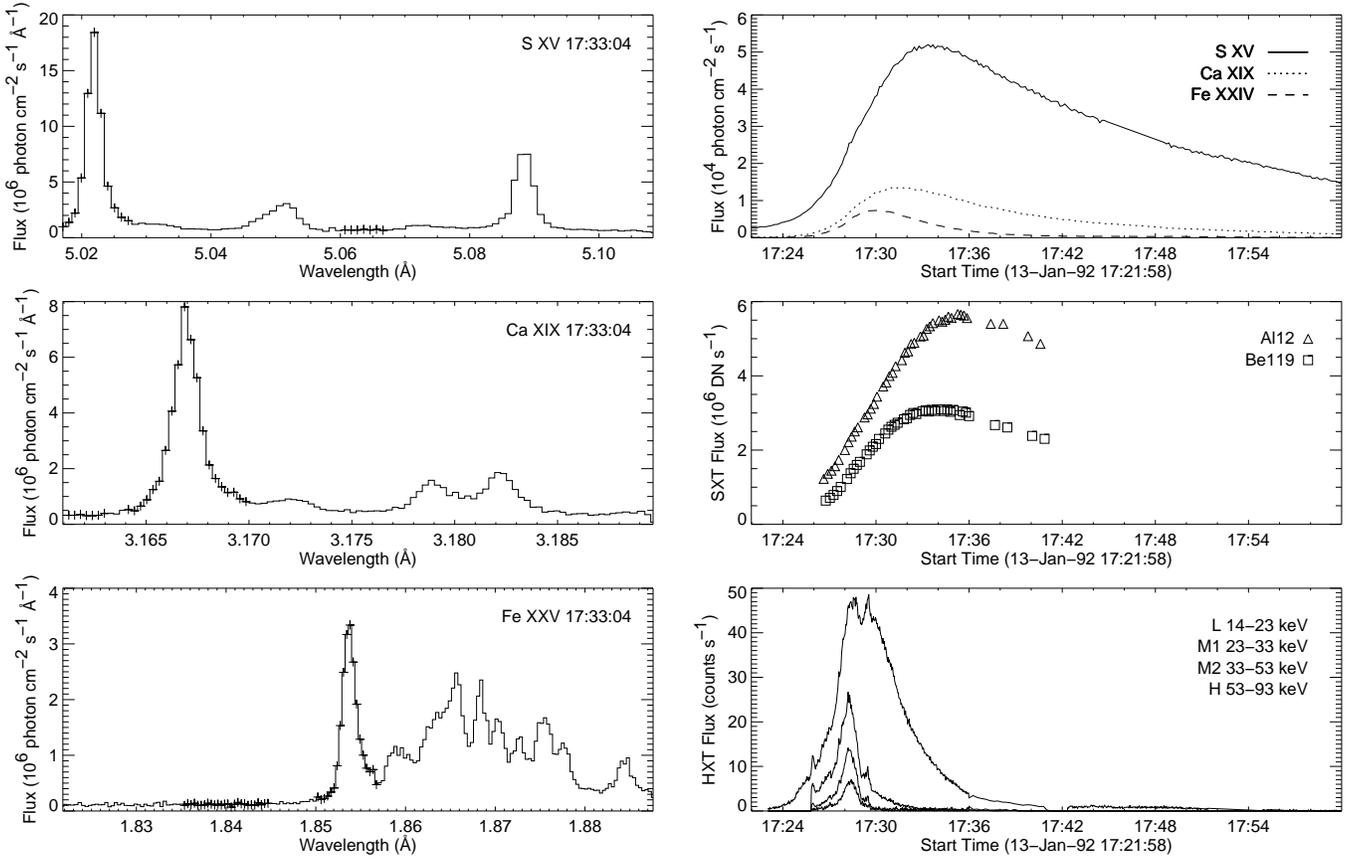}}
 \caption{\textit{Yohkoh} observations of the Mausa
 flare. (\textit{left panels}) BCS spectra in \ion{S}{15},
 \ion{Ca}{19}, \ion{Fe}{25}. The crosses indicate the spectral regions
 used to compute the intensities for the resonance line and the
 continuum. (\textit{right panels}) Light curves for BCS, SXT, and
 HXT.}
 \label{fig:lc}
 \end{figure*}

 SXT filter ratios suggest temperature in the range of 10--12\,MK in
 the brightest regions of the arcade. SXT also indicates a region of
 elevated temperatures ($\sim20$\,MK) in the faint region slightly
 above the arcade (e.g., \citealt{doschek1995,tsuneta1997}). BCS
 spectra, such as those shown in Figures~\ref{fig:lc}, indicate peak
 temperatures ranging from about 17\,MK (\ion{S}{15}) to 20\,MK
 (\ion{Fe}{25}). The differences between the temperatures derived from
 SXT and BCS suggest that the flare arcade is multi-thermal
 \citep{doschek1995}. Using the BCS fluxes in combination with
 spatially integrated SXT intensities yields emission measure
 distributions consistent with multi-thermal plasma
 \citep{mctiernan1999}.

\section{Single Loop Hydrodynamic Simulations}

 To illustrate the difficulties with previous solar flare hydrodynamic
 simulations we consider the modeling of the Masuda event as a single
 loop. We simulate the flare using the Naval Research
 Laboratory solar flux tube model (e.g., \citealt{mariska1987}) and
 the electron beam formalism of \cite{mariska1989}. The injected
 electron spectrum is assumed to be of the form
 \begin{equation}
    F_{in}(E_0,t) = \frac{4(\delta-2)}{\delta+2}
                 \frac{F_{max}g(t)}{E_c^2}
                 \left\{%
               \begin{array}{ll}
         \left(\frac{E_0}{E_c}\right)^2,         & \mbox{$E_0\le E_c$}, \\ 
         \left(\frac{E_0}{E_c}\right)^{-\delta}, & \mbox{$E_0\ge E_c$}.
               \end{array} \right.
 \label{eq:beam}
 \end{equation}
 Here $E_c$ is the low-energy cutoff, $F_{max}$ is the energy flux
 injected into the loop, and $g(t)$ is the temporal envelope on the
 heating. For thick target bremsstrahlung $\delta$ is related to the
 spectral index ($\gamma$) of the observed non-thermal emission by
 $\delta=\gamma+1$ \citep{emslie1988}. We use the loop length inferred
 from the observations and a spectral index of $\delta=4$ to
 approximate the observations.  Note that the energy deposition can be
 expressed as an analytic function of the integrated column density
 when $\delta$ is an even positive integer. We use a triangular
 envelope with a width of 80\,s that fits, at least approximately, the
 observed hard X-ray profile at the higher energies.

\begin{figure}[t!]
 \centerline{%
 \includegraphics[scale=0.585]{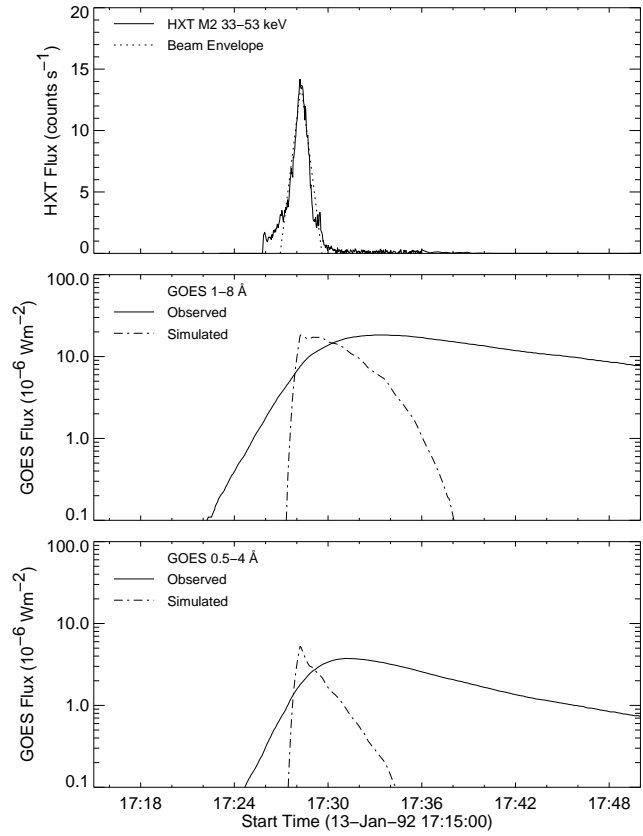}}
 \caption{Single loop hydrodynamic modeling of the Masuda
 event. (\textit{top panel}) The observed HXT hard X-ray light curve
 from 33--53\,keV and the assumed envelope on the beam
 heating. (\textit{bottom panels}) The observed and simulated GOES
 soft X-ray light curves.}
 \label{fig:sim1}
 \end{figure}

 The only adjustable parameters in the simulation are the low-energy
 cutoff, the energy flux injected into the loop, and the
 cross-sectional area of the loop. For a fixed value of $E_c$ the
 injected energy flux and the loop area are constrained by the need to
 reproduce the magnitude of both \textit{GOES} soft energy fluxes. In
 Figure~\ref{fig:sim1} we show the results from a simulation with
 $E_c=10$\,keV, $F_{max}=3\times10^{10}$\,ergs cm$^{-2}$, and a total
 volume of $2.8\times10^{27}$\,cm$^3$. The densities and temperatures
 computed from the simulation have been convolved with the GOES
 temperature response to compute the expected emission in each channel
 as a function of time. From this comparison it is clear that a single
 loop model will have a difficult time reproducing the observed
 emission. In the observations there is a delay of about 240\,s
 between the peak of the hard X-rays and the peak of the soft X-ray
 emission. The observations also show a very slow decline in the soft
 X-ray emission.  The simulated light curves, in contrast, show both a
 rapid increase and a rapid decay, suggesting that the soft X-ray
 emission is very sensitive to the energy deposition.

 It is tempting to posit that modifications to the heating function
 assumed in this exercise would improve the agreement between the
 observations and the numerical simulation. Variations in the spectral
 index or in the low energy cutoff, for example, would change the
 observed light curves. It is also possible to abandon the requirement
 that the heating be solely due to precipitating electrons and
 superimpose a second, more slowly varying heating on the loop.  In
 particular, several previous studies have considered the role of
 extended heating in explaining the decay of the soft X-ray emission
 (e.g., \citealt{reale1997}). It is clear from the observations,
 however, that extended heating during the decay phase cannot be the
 primary cause of the extended life-time of the soft X-ray
 emission. In this event, for example, the first H$\alpha$ post-flare
 loops appear at 17:38 UT \citep{hwang1995} when, as is indicated by
 the \textit{GOES} light curves and the SXT images, the high
 temperature soft X-ray emission is still substantial. Furthermore,
 these H$\alpha$ loops appear at heights below the soft X-ray loops
 observed with SXT. It is unambiguous that single loop modeling cannot
 account for the observed evolution of the flare arcade. This
 conclusion has been reached by other authors (e.g.,
 \citealt{moore1980,svestka1982, schmieder1995}). This understanding,
 however, has not been widely incorporated into the hydrodynamic
 modeling of solar flares.

\section{Multi-Thread Hydrodynamic Modeling} \label{sec:modeling}

 One impediment to modeling a solar flare as the succession of
 independently heated threads is the difficulty in determining the
 heating parameters for each thread. The observables from a flare,
 such as the \textit{GOES} soft X-ray fluxes, are typically used to
 compute physical properties of the flare arcade, such as the
 temperature and emission measure. These quantities are difficult to
 use as inputs to a hydrodynamic simulation.

 Recent work by \cite{warren2004a} has investigated the relationship
 between the \textit{GOES} soft X-ray fluxes and the parameters
 relevant to hydrodynamic simulations, such as the total energy
 deposited into a thread and the volume of the thread. They found that
 the peak fluxes in the \textit{GOES} channels were related to the
 energy input into the loop by 
 \begin{equation}
   F_{1-8}(t_P) \simeq 3.68\times10^{-35}\left[\frac{EL}{V}\right]^{1.75}
   \frac{V}{L^2}
   \label{eq:flong}
 \end{equation}
 and
 \begin{equation}
   F_{0.5-4}(t_P) \simeq 4.42\times10^{-42}\left[\frac{EL}{V}\right]^{2.24}
   \frac{V}{L^2}.
   \label{eq:fshort}
 \end{equation}
 Here $E$ is the total energy deposited in the loop, $L$ is the total
 loop length, and $V$ is the loop volume. The cross-sectional area
 ($A$) is $V/L$. Conceptually, we can think of the hydrodynamic loop
 equations as describing the evolution of the plasma in a thin,
 semi-annular volume where the loop length corresponds to the
 semi-diameter. 
 
 Assuming that the loop length can be determined independently,
 equations~\ref{eq:flong} and \ref{eq:fshort} can be inverted to
 express the energy deposition ($E$) and loop volume ($V$) in terms of
 the peak \textit{GOES} soft X-ray fluxes. These expressions, however,
 describe the relationship between $E$, $V$, and the \textit{GOES}
 soft X-ray fluxes for a single thread. In our multi-thread simulation
 we model the flare as a succession of independently heated
 threads. To account for the contribution of threads that have been
 heated previously and are cooling we compute the ``residual''
 \textit{GOES} flux for each channel,
 \begin{equation}
 \Delta F^i(t_p) = \left[ F_{obs}(t+t_c) - \sum_{j=1}^{i-1}F_{sim}^j(t+t_c)
   \right].
 \end{equation}
 Here $t_c$ is essentially the conductive cooling time (the time at
 which the emission measure of the thread will reach its maximum
 value). This offset reflects the fact that there will be a delay
 between the introduction of a thread and the time when it reaches its
 maximum emission in soft X-rays. In the model $t_c$ is estimated from
 the simulation of the previous thread. For each thread after the
 first, the residual \textit{GOES} fluxes are used to compute the
 energy and volume for the thread.

 Computing a multi-thread simulation of a solar flare involves the
 following steps
 \begin{enumerate}
 \item Using background subtracted \textit{GOES} fluxes we compute the
   energy and volume needed to reproduce the observed soft X-ray
   emission.
 \item We perform a time-dependent hydrodynamic simulation using the
   energy derived from the \textit{GOES} data.
 \item Calculate the evolution of the \textit{GOES} fluxes from the
   hydrodynamic simulation and the thread volume derived from the
   observations.
 \item Compute the residual \textit{GOES} flux in each channel, which
   are then used in step 1 for the next thread in the simulation
 \end{enumerate}
 In principal, the scaling laws presented in Equations~\ref{eq:flong}
 and \ref{eq:fshort} can be used to approximate the input energy and
 volume for a thread from the observed \textit{GOES} soft X-ray
 fluxes.  However, these scaling laws are not useful for computing the
 temporal evolution of a thread and cannot be used to determine the
 residual fluxes that are important in the determining the parameters
 for the next thread in the simulation. Thus the use of the scaling
 laws necessitates that the simulations must be run in series, making
 the process of simulating a flare very time consuming.

\begin{figure}[t!]
 \centerline{%
 \includegraphics[scale=0.585]{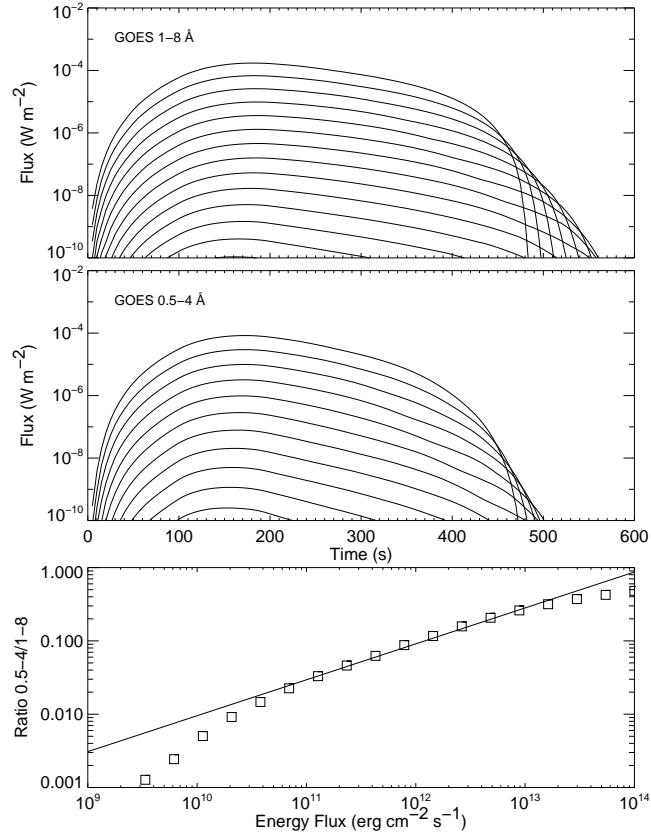}}
 \caption{(\textit{top panels}) Simulated \textit{GOES} light curves
 as a function of time and input energy flux.  (\textit{bottom panel})
 The ratio of the 0.5--4 to 1--8\,\AA\ \textit{GOES} channels. The
 loop length has been fixed at approximately 39\,Mm in these
 simulations. An area of $10^{16}$\,cm$^2$ has been assumed. The solid
 line represents the theoretical ratio derived from the scaling
 laws.}
 \label{fig:grid}
 \end{figure}

 To allow for parallel processing we compute a series of full
 hydrodynamic simulations using a wide range of input energies. For
 each of these simulations we calculate the evolution of the
 \textit{GOES} fluxes in both channels. An illustration of this
 calculation is shown in Figure~\ref{fig:grid}. With this ``grid'' of
 solutions we can interpolate to find the energy and volume needed to
 reproduce the observed \textit{GOES} fluxes. Additionally we can
 interpolate in time to estimate the evolution of the thread as it
 would be observed with \textit{GOES}. The use of interpolated light
 curves allows for the hydrodynamic simulation parameters for all of
 the threads to be determined very rapidly. Once the simulation
 parameters have been determined from the interpolated light curves
 the full hydrodynamic simulations can be performed in parallel,
 leading to a dramatic increase in computational efficiency.

 Since the scaling laws incorporate significant approximations that
 are not present in the solutions to the hydrodynamic equations, the
 use of interpolated light curves makes the flare simulations
 considerably more accurate. It is also important to note that the
 scaling laws have not been throughly compared with results from
 hydrodynamic simulations. \cite{warren2004a} only considered
 variations in the input energy for a single loop length. As
 illustrated in Figure~\ref{fig:grid}, the scaling laws appear to work
 well for very short loops. Longer loops are likely to be more
 problematic. More exhaustive comparisons are currently in progress
 and will be reported in a future publication.

 For this work we make two simplifying assumptions. First, we assume
 that the threads are heated directly and that energy transport during
 the initial phase of the flare is dominated by thermal conduction.
 The functional form of the heating is assumed to be
 \begin{equation}
   E_H(s,t) = E_0 +
   g(t)E_{Flare}\exp\left[-\frac{(s-s_0)^2}{2\sigma_s^2}\right],
   \label{eq:direct}
 \end{equation}
 where $s_0$ designates the location of the impulsive heating,
 $\sigma_s$ is the spatial width of the heating, and $E_{Flare}$ is a
 constant that determines the maximum amplitude of the heating. The
 background heating parameter, $E_0$, is chosen so that the
 equilibrium atmosphere is very cool ($\sim0.5$\,MK) and tenuous and
 has little effect on the evolution of the loop. A more realistic
 treatment of the energy deposition would include the contribution of
 energetic particles precipitating into the chromosphere.
 Figure~\ref{fig:sim1}, however, suggests that energetic particle
 precipitation is responsible for only a small fraction of the
 observed \textit{GOES} emission during this flare.

 The second simplifying assumption is that a fixed loop length can be
 used for all of the hydrodynamic simulations.  The SXT observations
 indicate that the height of the arcade increases by about 25\% during
 the time that SXT images are available (17:28--17:40 UT). Changes in
 the loop length of this magnitude do not dramatically alter the
 evolution of a thread. Incorporating changes in the loop length would
 require the calculation of a 2 dimensional grid of solutions where
 both the energy and the loop length are varied. This would add
 significant computation to the simulation. Also, since there are SXT
 images only near the peak of the event we have no way to estimate the
 variation in loop length during most of the decay.  For the
 simulations of this flare we fix the loop length for each thread at
 39\,Mm. All of the simulations shown in Figure~\ref{fig:grid} are
 computed with this loop length and the heating function described by
 Equation~\ref{eq:direct}.

\begin{figure*}[t!]
 \centerline{%
 \includegraphics[scale=0.585,angle=90]{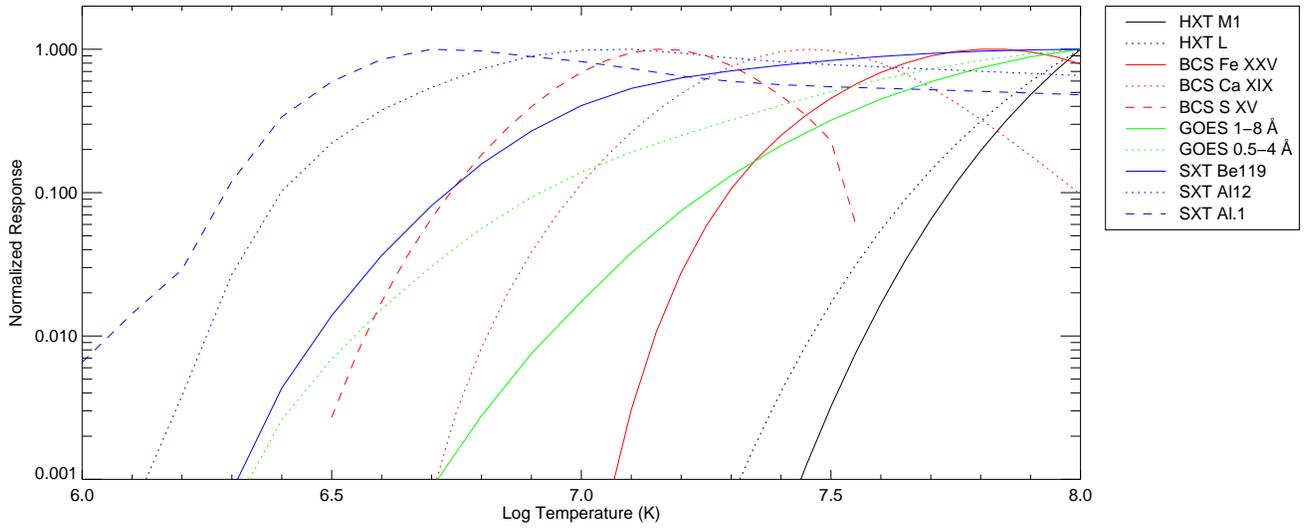}}
 \caption{Temperature response curves the for SXT Al.1, Al12, and
 Be119 filters, \textit{GOES} 0.5--4 and 1--8\,\AA\ channels, the BCS
 \ion{S}{15}, \ion{Ca}{19}, and \ion{Fe}{25} resonance lines, and the
 HXT L and M1 channels. The response in each channel has been
 normalized to its maximum value.}
 \label{fig:resp}
 \end{figure*}

 The hydrodynamic simulations yield the temperature and density along
 the length of each thread in the arcade. This allows us to compute
 the emission for any wavelength range in the solar spectrum that is
 dominated by optically thin emission. For this paper we focus on
 computing light curves for the HXT, BCS, and SXT instruments on
 \textit{Yohkoh}. For SXT and HXT we use the standard \verb+SolarSoft+
 routines \verb+SXT_FLUX+ and \verb+HXT_THCOMP+ to compute the
 response as a function of temperature. For BCS we compute synthetic
 spectra over a wide range of temperatures using \verb+BCS_SPEC+ and
 sum over the resonance line as indicated in Figure~\ref{fig:lc} to
 determine the response as a function of temperature. The response
 curves shown in Figure~\ref{fig:resp} illustrate how these
 instruments span the temperature range from approximately 2\,MK to
 above 100\,MK.

 It is encouraging to note that many of the significant parameters in
 the multi-thread simulation are derived from the observations and
 are not adjustable. The energy deposited into each thread and the
 volume of each thread are inferred from the observed \textit{GOES}
 fluxes. The loop length can generally be determined from image data,
 although projection effects can be difficult to account for. There
 are, however, several parameters in the simulations that are
 potentially unconstrained by the \textit{GOES} observations, such as
 the rate at which threads are introduced into the flare arcade.
 Similarly, the details of the energy deposition, such as its
 magnitude, duration, and spatial scale, are also largely
 unconstrained by the \textit{GOES} observations. The peak fluxes are
 largely determined by the total energy deposited in the loop (e.g.,
 \citealt{winebarger2004}). Additionally, it should not be forgotten
 that there are a number of simplifying assumptions incorporated into
 the hydrodynamic simulations. The model chromosphere used in the
 hydrodynamic code, in particular, is highly simplified.

\section{Multi-Thread Hydrodynamic Simulations}

\begin{figure*}[t!]
 \centerline{%
 \includegraphics[scale=0.585,angle=90]{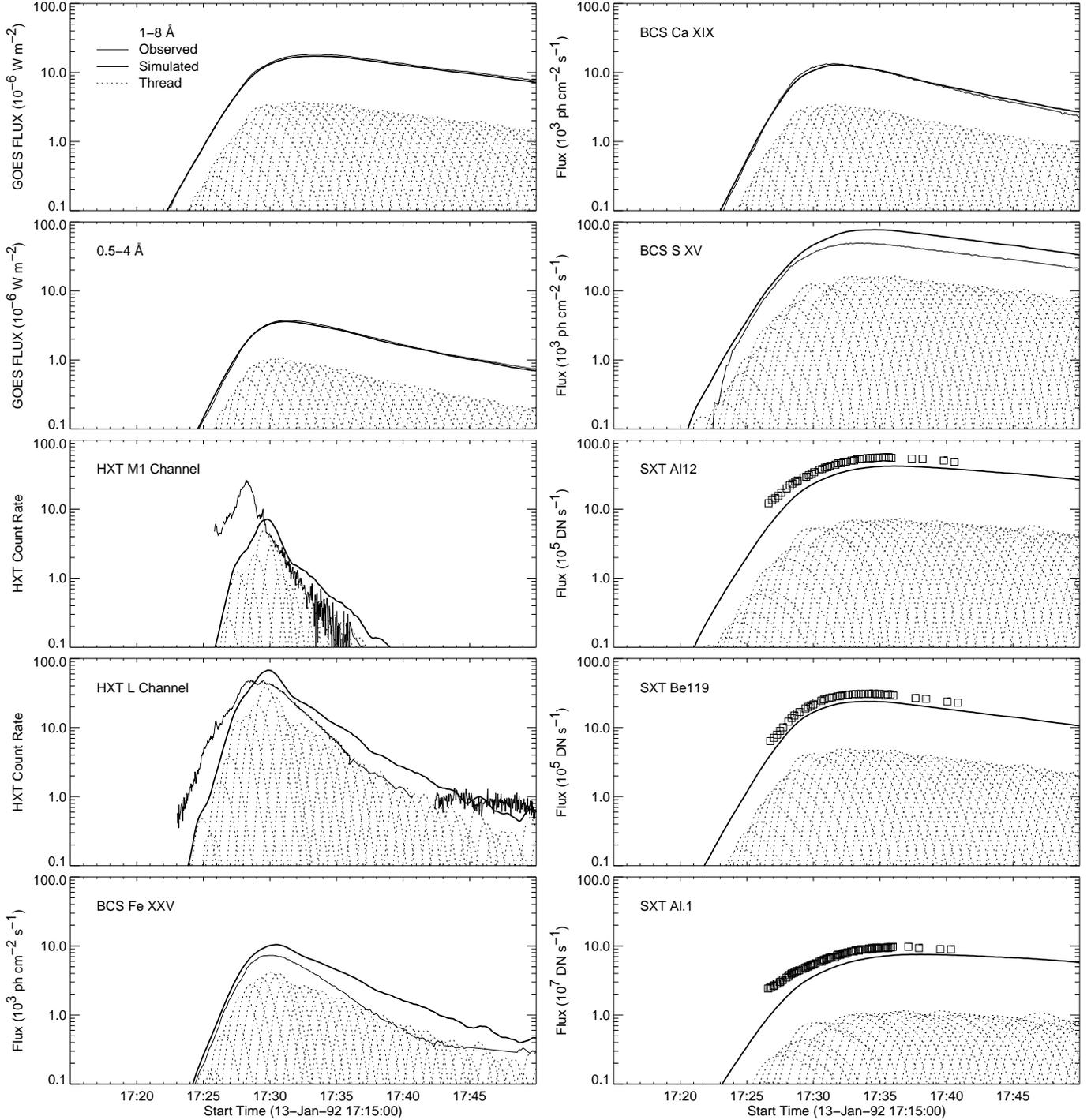}}
 \caption{Observed and simulated light curves for the Masuda flare of
 1992 January 13. No scaling factors have been applied to any of the
 simulated light curves. In each panel the thick solid line represents
 simulated light curve computed by summing the contribution of each
 thread. The light curves for the individual threads are indicated
 by the dotted lines. The observed fluxes are indicated by a thin
 solid line (\textit{GOES}, HXT, and BCS) or by squares (SXT). See the
 text for a discussion of background subtraction for the observed
 fluxes.}
 \label{fig:simm}
 \end{figure*}

 An example of a multi-thread simulation for the Masuda flare is shown
 in Figure~\ref{fig:simm}. In this simulation we have used 50 threads
 each introduced 40\,s apart. For each thread the hydrodynamic code
 was run using the heating function given in Equation~\ref{eq:direct}
 with $\sigma_H=10^8$\,cm, $s_0$ set equal to the loop apex, and with
 $g(t)$ a triangular envelop with a width of 200\,s. As noted
 previously we have fixed the loop length at 39\,Mm and we focus our
 attention on the rise phase and beginning of the decay of the flare.

 The results presented in Figure~\ref{fig:simm} indicate that the
 simulation reproduces the flux in each \textit{GOES} channel very
 well. The differences between the observed and simulated light curves
 are typically about 20\%. This good agreement is expected since the
 parameters for each thread were inferred from the \textit{GOES}
 soft X-ray measurements.  The observations from HXT, BCS, and SXT on
 \textit{Yohkoh} are not used to infer the simulation parameters and
 offer an independent assessment of the simulation results. The
 \textit{Yohkoh} instruments also cover a wider range of temperature
 than the two \textit{GOES} channels. The light curves shown in
 Figure~\ref{fig:simm} show that the simulation does a credible job of
 reproducing the observed \textit{Yohkoh} fluxes. In this flare the
 highest temperature emission (e.g., HXT L and BCS \ion{Fe}{25}) peaks
 earliest and decays quickly while the lower temperature emission
 (e.g, BCS \ion{S}{15} and SXT Al.1) peaks later and decays relatively
 slowly, a trend that is observed in many flares (e.g,
 \citealt{sterling1997}). The simulations do a particularly good job
 of reproducing this behavior. 

 The most significant discrepancy between the simulation and the
 observations is for the HXT L and M1 channels very early in the
 event. At these times the emission in these HXT channels is likely to
 be non-thermal bremsstrahlung produced by precipitating high energy
 electrons. At present this modeling only accounts for the thermal
 emission in a flare. Since the observed BCS \ion{Fe}{25} light curve
 is well matched by the simulation it does not appear that the
 discrepancies in the HXT light curves is caused by the simulation
 underestimating the temperature. 

 The modeled SXT light curves generally reproduce the temporal
 evolution of the observations. The observed SXT fluxes, however, are
 systematically higher than the simulated fluxes by about a factor of
 2. Since this discrepancy is systematic, it may be due to
 inconsistencies in the radiometric calibration between \textit{GOES}
 and SXT.

 Note that for these comparisons we have subtracted the first flux
 measurements from the BCS observations to account for the
 background. This correction is significant only for the BCS
 \ion{S}{15} emission. For HXT we have estimated the background using
 fluxes from a period late in the flare. The SXT data during this
 orbit cover only a small period near the peak of the flare and
 background subtraction is not possible, and this contributes to the
 discrepancies between the observation and simulation in these
 filters.

\begin{figure*}[t!]
 \centerline{%
 \includegraphics[scale=0.585,angle=90]{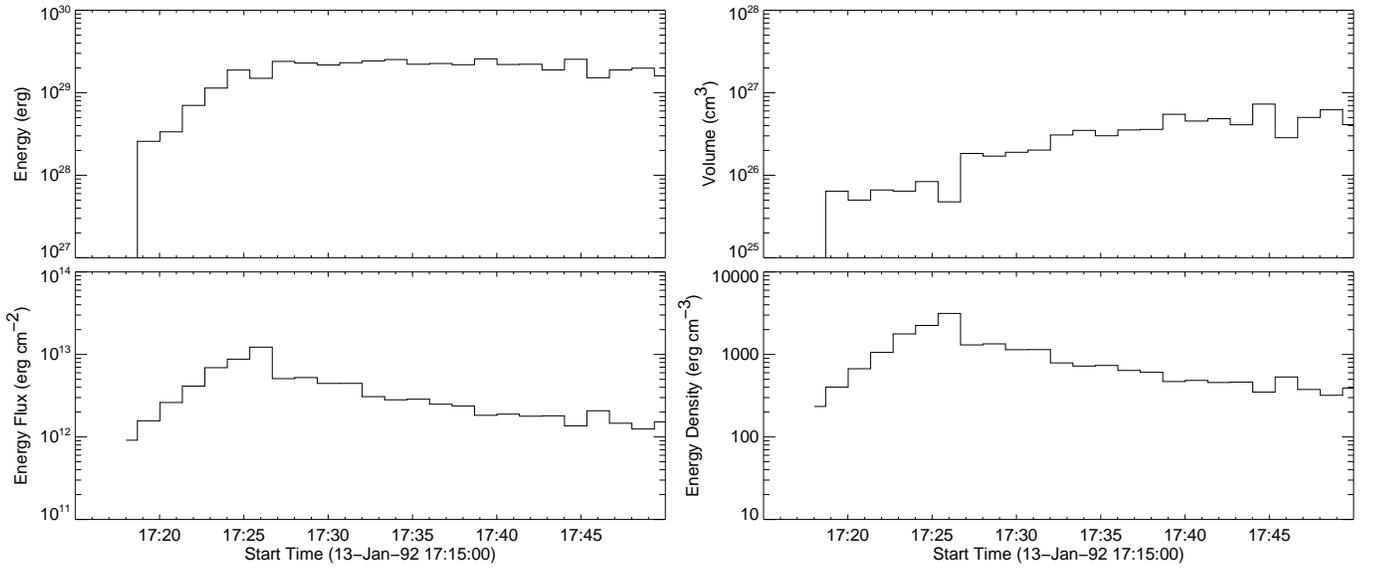}}
 \caption{Parameters for each thread in the simulation as a function
 of time. Shown are the total energy, volume, average energy flux
 ($E/A$), and average energy density ($E/V$) for each thread. Note
 that the times correspond to the beginning of the heating for each
 thread.}
 \label{fig:energy}
 \end{figure*}

 The energies and volumes for each thread in the simulation are shown
 in Figure~\ref{fig:energy}. Also shown are the average energy flux
 ($E/A$) and the energy density ($E/V$) for each thread. These
 quantities are computed by dividing the total energy input into the
 thread by the area or the volume of the thread.  The equations
 presented in Section~\ref{sec:modeling} suggest that the peak
 temperatures and densities in each thread are largely determined by
 the energy flux. Thus the largest energy fluxes occur early in the
 flare and generally correspond to those threads which have the
 highest peak temperatures and densities. The maximum \textit{GOES}
 ratio of 0.25 occurs at about 17:29 UT, close to the peaks in the HXT
 L and BCS \ion{Fe}{25} emission.  After this time the flux in the
 0.5--4\,\AA\ channel declines more rapidly than the flux in the
 1--8\,\AA\ channel, suggesting declining peak temperatures and
 therefore declining energy fluxes. Solving Equations
 Equations~\ref{eq:flong} and \ref{eq:fshort} for the energy flux
 yields
 \begin{equation}
   \frac{EL}{V}\sim
   \left[\frac{F_{0.5-4}(t_p)}{F_{1-8}(t_p)}\right]^{2.04}.
   \label{eq:ratio}
 \end{equation}
 Since the \textit{GOES} fluxes decay exponentially (at least
 approximately) the energy flux should also decay exponentially during
 the decay of the flare. This behavior is evident in the simulation
 results.

 The plot of the total energy deposited into each thread shown in
 Figure~\ref{fig:energy} is rather surprising. The input energy
 increases during the rise phase phase of the flare, as
 expected. During the decay, however, the input energy remains
 relatively constant. Given the exponential decay in both the
 \textit{GOES} emission and the energy flux it seems reasonable to
 assume that the total energy input into each tread would also decline
 rapidly during the decay phase.  However, Equations~\ref{eq:flong}
 and \ref{eq:fshort} indicate that the \textit{GOES} intensities
 should decay even faster than the energy flux, since the intensities
 are, approximately, quadratic functions of the energy flux. The
 \textit{GOES} 0.5--4 to 1--8\,\AA\ ratio, which is broadly indicative
 of the temperature, goes as the square root of the energy flux (see
 Equation~\ref{eq:ratio}) and should therefore decay more slowly than
 the \textit{GOES} intensities. This is the opposite of what is
 typically observed. Typically the temperature declines more rapidly
 than the \textit{GOES} intensities (e.g.,
 \citealt{sterling1997}). Thus the rise in the thread volume counters,
 at least partially, the rapid decline in the energy flux and leads to
 a much slower decline in the total energy during the decay of the
 flare than anticipated.

 Intuitively, the rise in the thread volume is largely the result of
 the rapid rise of the reconnection region and the corresponding
 increase in the thread length.  In this simulation, however, the
 length of the threads has remained fixed. The $L^{-1}$ scaling of the
 \textit{GOES} flux Equations~\ref{eq:flong} and \ref{eq:fshort}
 suggest that increasing the thread length will actually accelerate
 the decline in the \textit{GOES} fluxes and lead to an even more
 rapid increase in the thread volume.  However, since the simulation
 parameters are computed from the residual \textit{GOES} fluxes, the
 cooling of the previous threads plays an important role in
 determining the flare energy. As the thread length increases the
 conductive cooling time increases and the cooling of the thread
 proceeds more slowly (e.g., \citealt{cargill1995}). This leads to
 smaller residual fluxes and potentially smaller volumes.

 The nonlinearity of the flare simulation makes it difficult to
 determine exactly how the energy will vary when the length is
 properly accounted for. The rapid decline of the highest temperature
 emission, as indicated by the HXT L and BCS \ion{Fe}{25} light
 curves, suggests that the energy flux declines rapidly during the
 decay. Given the strong dependence of the \textit{GOES} fluxes on the
 energy flux it is clear that the volume must rise rapidly during the
 decay of the flare. Thus it seems highly likely that the energy
 released during the decay phase is substantial, as indicated by these
 simulation results.
 
 In this simulation assumptions have been made regarding the location
 and duration of the heating. We have also made assumptions about the
 rate at which new threads are introduced into the simulation. Recent
 work on impulsive heating in hydrodynamic simulations suggest that
 variations in the duration of the heating are the most likely to have
 the most significant observable consequences. The simulation
 algorithm yields the total energy and volume for each tread, it says
 nothing about the rate at which energy is released into the
 thread. \cite{winebarger2004} have shown that once the thread reaches
 the radiative phase of the cooling the evolution of the thread is
 determined largely by the total energy. During the conductive phase,
 in contrast, the details of the energy release, such as the heating
 rate, do influence the evolution of the density and temperature. The
 peak temperature is particularly sensitive to how impulsive the
 heating is. Since the HXT channels and the BCS \ion{Fe}{25} line are
 very sensitive to the presence of very high temperature plasma we
 expect these light curves to discriminate between heating parameters.

\begin{figure}[t!]
 \centerline{%
 \includegraphics[scale=0.585,angle=90]{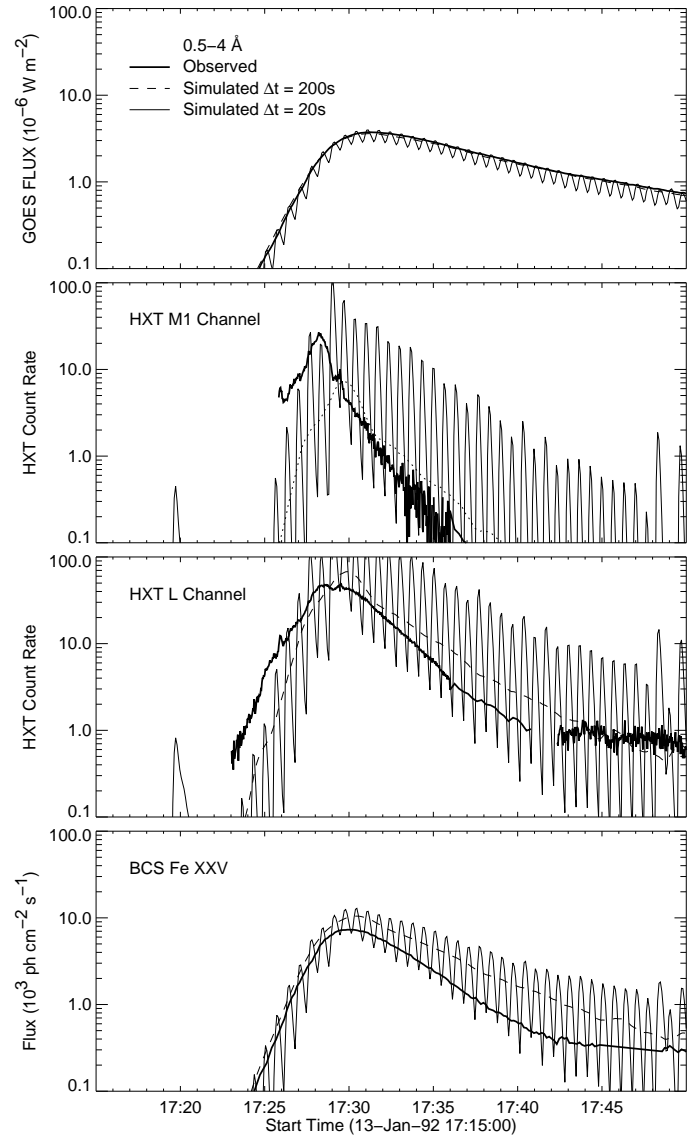}}
 \caption{Light curves for emission formed at high temperatures for
 relatively gentle heating (200\,s) and impulsive heating (20\,s).}
 \label{fig:simm2}
 \end{figure}

 To investigate this we compute another simulation where the width of
 the heating envelope has been set to 20\,s. For this case a grid of
 solutions to the hydrodynamic equations are computed using this
 heating profile and the flare simulation is performed using the
 procedure outlined previously. The resulting light curves for
 \textit{GOES} 0.5--4\,\AA, HXT M1, HXT L, and BCS \ion{Fe}{24} are
 shown in Figure~\ref{fig:simm2}. The more impulsive nature of the
 heating in this simulation leads to higher maximum temperatures for
 each thread. The higher temperatures are reflected in the light
 curves for the highest temperature emission. The HXT M1 channel shows
 greatly enhanced emission for the more impulsive heating. The impact
 of the shortened heating time scale becomes progressively smaller for
 emission formed at lower temperatures. In Figure~\ref{fig:simm2} we
 see that the simulation with the more gentle heating ($200$\,s) is in
 better agreement with the observations than the simulation with the
 impulsive heating.

\section{Discussion}

 One dimensional hydrodynamic modeling represents an important link
 between our physical understanding of processes in the solar corona
 and solar observations.  The details of the energy release through
 magnetic reconnection, for example, can by investigated with
 magnetohydrodynamics (MHD), but three dimensional MHD simulations
 generally lack the spatial resolution needed to properly track the
 flow of mass and energy through the solar atmosphere. This limits our
 ability to directly compare MHD simulation results with observation.

 We have shown that it is possible to use hydrodynamic simulations to
 reproduce many of the salient properties of the observed light
 curves. Though the heating function assumed in the hydrodynamic
 simulation is phenomenological, these simulations do provide
 important constraints on the energy release mechanism. These
 simulations of the Masuda flare, for example, indicate that very
 impulsive heating with short heating time scales ($\sim20$\,s) would
 lead to the formation of very high temperature plasma. The
 \textit{Yohkoh} HXT observations provide an upper bound on the plasma
 temperature in the bulk of the arcade and suggest that such very
 impulsive heating is not consistent with the available data. The
 simulations that are more broadly consistent with the observations
 have a heating time scale that is relatively long ($\sim200$\,s), and
 any successful model of magnetic reconnection must reproduce
 this. Similarly, these simulations suggest the the rate of energy
 release decays slowly during the decay phase of the flare. This is
 potentially another important constraint on the energy release
 mechanism.

 In addition to providing a link between the observations and the
 details of the energy release during a flare, these simulations also
 have practical applications. The changes in the solar soft X-ray and
 EUV irradiance during a flare, for example, perturb the Earth's
 ionosphere and provide an ideal way to test our understanding of
 physical processes in this region of the upper atmosphere (e.g.,
 \citealt{meier2002}). Spectrally resolved EUV irradiance observations
 taken during solar flares are rare, however. Our simulation results
 provide a means for computing time-dependent soft X-ray and EUV
 irradiance variations associated with a flare that can be used in
 modeling the ionospheric response. The initial application of these
 modeling techniques have provided encouraging results
 \citep{huba2005}.

 In performing these flare simulations we have made several
 simplifying assumptions that reduce the computational complexity of
 the calculations. For example, we have assumed a constant loop length
 for all of the hydrodynamic simulations. This allows us to use a
 single grid of solutions to determine the energy and volume for each
 thread.  Accommodating a varying loop length in the algorithm will
 not be difficult, but will increase the cpu time required for a
 simulation substantially. We have also assumed that energy transport
 during the initial phase of the flare is solely due to thermal
 conduction. In reality, bursts of hard X-rays are present during the
 rise phase of almost every large flare. Accounting for these hard
 X-ray bursts can be accomplished by computing grid of solutions for
 beam heated threads and partitioning the energy between in situ
 heating and beam heating using the observed high energy hard X-ray
 emission. Such modeling would provide important constraints on the
 partition of thermal and non-thermal heating during a flare.
 

 \acknowledgments This research was supported by NASA's Sun-Earth
 Connection Guest Investigator program and the NRL/ONR 6.1 basic
 research program.



 \clearpage

 \clearpage

 \clearpage

 \clearpage

 \clearpage

 \clearpage

 \clearpage

 \clearpage

\begin{thebibliography}{}

\bibitem[\protect\citeauthoryear{{Aschwanden} \& {Alexander}}{{Aschwanden} \&
  {Alexander}}{2001}]{aschwanden2001c}
{Aschwanden}, M.~J.,  \& {Alexander}, D. 2001, \solphys, 204, 91

\bibitem[\protect\citeauthoryear{{Aschwanden} et~al.}{{Aschwanden}
  et~al.}{1996}]{aschwanden1996}
{Aschwanden}, M.~J., {Hudson}, H., {Kosugi}, T.,  \& {Schwartz}, R.~A. 1996,
  \apj, 464, 985

\bibitem[\protect\citeauthoryear{{Cargill}, {Mariska}, \&
  {Antiochos}}{{Cargill} et~al.}{1995}]{cargill1995}
{Cargill}, P.~J., {Mariska}, J.~T.,  \& {Antiochos}, S.~K. 1995, \apj, 439,
  1034

\bibitem[\protect\citeauthoryear{{Culhane} et~al.}{{Culhane}
  et~al.}{1991}]{culhane1991}
{Culhane}, J.~L., et~al. 1991, \solphys, 136, 89

\bibitem[\protect\citeauthoryear{{Doschek}, {Strong}, \& {Tsuneta}}{{Doschek}
  et~al.}{1995}]{doschek1995}
{Doschek}, G.~A., {Strong}, K.~T.,  \& {Tsuneta}, S. 1995, \apj, 440, 370

\bibitem[\protect\citeauthoryear{{Hori} et~al.}{{Hori} et~al.}{1997}]{hori1997}
{Hori}, K., {Yokoyama}, T., {Kosugi}, T.,  \& {Shibata}, K. 1997, \apj, 489,
  426

\bibitem[\protect\citeauthoryear{{Hori} et~al.}{{Hori} et~al.}{1998}]{hori1998}
{Hori}, K., {Yokoyama}, T., {Kosugi}, T.,  \& {Shibata}, K. 1998, \apj, 500,
  492

\bibitem[\protect\citeauthoryear{{Huba} et~al.}{{Huba} et~al.}{2005}]{huba2005}
{Huba}, J.~D., {Warren}, H.~P., {Joyce}, G., {Pi}, X., {Iijima}, B.,  \&
  {Coker}, C. 2005, \grl, in press

\bibitem[\protect\citeauthoryear{{Kosugi} et~al.}{{Kosugi}
  et~al.}{1991}]{kosugi1991}
{Kosugi}, T., et~al. 1991, \solphys, 136, 17

\bibitem[\protect\citeauthoryear{{Mariska}}{{Mariska}}{1987}]{mariska1987}
{Mariska}, J.~T. 1987, \apj, 319, 465

\bibitem[\protect\citeauthoryear{{Mariska}, {Doschek}, \& {Bentley}}{{Mariska}
  et~al.}{1993}]{mariska1993}
{Mariska}, J.~T., {Doschek}, G.~A.,  \& {Bentley}, R.~D. 1993, Astrophys. J.,
  419, 418

\bibitem[\protect\citeauthoryear{{Mariska}, {Emslie}, \& {Li}}{{Mariska}
  et~al.}{1989}]{mariska1989}
{Mariska}, J.~T., {Emslie}, A.~G.,  \& {Li}, P. 1989, \apj, 341, 1067

\bibitem[\protect\citeauthoryear{{Mariska} \& {Zarro}}{{Mariska} \&
  {Zarro}}{1991}]{mariska1991}
{Mariska}, J.~T.,  \& {Zarro}, D.~M. 1991, \apj, 381, 572

\bibitem[\protect\citeauthoryear{{Masuda} et~al.}{{Masuda}
  et~al.}{1995}]{masuda1995}
{Masuda}, S., {Kosugi}, T., {Hara}, H., {Sakao}, T., {Shibata}, K.,  \&
  {Tsuneta}, S. 1995, \pasj, 47, 677

\bibitem[\protect\citeauthoryear{{McTiernan}, {Fisher}, \& {Li}}{{McTiernan}
  et~al.}{1999}]{mctiernan1999}
{McTiernan}, J.~M., {Fisher}, G.~H.,  \& {Li}, P. 1999, \apj, 514, 472

\bibitem[\protect\citeauthoryear{Meier et~al.}{Meier et~al.}{2002}]{meier2002}
Meier, R.~R., et~al. 2002, Geophys. Res. Lett., 29, 1461

\bibitem[\protect\citeauthoryear{{Moore} \& {Labonte}}{{Moore} \&
  {Labonte}}{1980}]{moore1980}
{Moore}, R.~L.,  \& {Labonte}, B.~J. 1980, in IAU Symp. 91: Solar and
  Interplanetary Dynamics, Vol.~91, 207

\bibitem[\protect\citeauthoryear{{Peres} et~al.}{{Peres}
  et~al.}{1987}]{peres1987}
{Peres}, G., {Reale}, F., {Serio}, S.,  \& {Pallavicini}, R. 1987, \apj, 312,
  895

\bibitem[\protect\citeauthoryear{{Reale} et~al.}{{Reale}
  et~al.}{1997}]{reale1997}
{Reale}, F., {Betta}, R., {Peres}, G., {Serio}, S.,  \& {McTiernan}, J. 1997,
  \aap, 325, 782

\bibitem[\protect\citeauthoryear{{Reeves} \& {Warren}}{{Reeves} \&
  {Warren}}{2002}]{reeves2002a}
{Reeves}, K.~K.,  \& {Warren}, H.~P. 2002, \apj, 578, 590

\bibitem[\protect\citeauthoryear{{Schmieder} et~al.}{{Schmieder}
  et~al.}{1995}]{schmieder1995}
{Schmieder}, B., {Heinzel}, P., {Wiik}, J.~E., {Lemen}, J., {Anwar}, B.,
  {Kotrc}, P.,  \& {Hiei}, E. 1995, \solphys, 156, 337

\bibitem[\protect\citeauthoryear{{Sterling} et~al.}{{Sterling}
  et~al.}{1997}]{sterling1997}
{Sterling}, A.~C., {Hudson}, H.~S., {Lemen}, J.~R.,  \& {Zarro}, D.~A. 1997,
  \apjs, 110, 115

\bibitem[\protect\citeauthoryear{{Svestka} et~al.}{{Svestka}
  et~al.}{1982}]{svestka1982}
{Svestka}, Z., {Dodson-Prince}, H.~W., {Mohler}, O.~C., {Martin}, S.~F.,
  {Moore}, R.~L., {Nolte}, J.~T.,  \& {Petrasso}, R.~D. 1982, \solphys, 78, 271

\bibitem[\protect\citeauthoryear{{Tandberg-Hanssen} \&
  {Emslie}}{{Tandberg-Hanssen} \& {Emslie}}{1988}]{emslie1988}
{Tandberg-Hanssen}, E.,  \& {Emslie}, A.~G. 1988, {The Physics of Solar Flares}
  (Cambridge and New York, Cambridge University Press, 1988, 286 p.)

\bibitem[\protect\citeauthoryear{Tsuneta et~al.}{Tsuneta
  et~al.}{1991}]{tsuneta1991}
Tsuneta, S., et~al. 1991, Solar Phys., 136, 37

\bibitem[\protect\citeauthoryear{{Tsuneta} et~al.}{{Tsuneta}
  et~al.}{1997}]{tsuneta1997}
{Tsuneta}, S., {Masuda}, S., {Kosugi}, T.,  \& {Sato}, J. 1997, \apj, 478, 787

\bibitem[\protect\citeauthoryear{{Wang} et~al.}{{Wang}
  et~al.}{1995}]{hwang1995}
{Wang}, H., {Gary}, D.~E., {Zirin}, H., {Kosugi}, T., {Schwartz}, R.~A.,  \&
  {Linford}, G. 1995, \apjl, 444, L115

\bibitem[\protect\citeauthoryear{Warren}{Warren}{2000}]{warren2000b}
Warren, H.~P. 2000, \apj, 536, L105

\bibitem[\protect\citeauthoryear{{Warren} \& {Antiochos}}{{Warren} \&
  {Antiochos}}{2004}]{warren2004a}
{Warren}, H.~P.,  \& {Antiochos}, S.~K. 2004, \apjl, 611, L49

\bibitem[\protect\citeauthoryear{Warren et~al.}{Warren
  et~al.}{1999}]{warren1999b}
Warren, H.~P., Bookbinder, J.~A., Forbes, T.~G., Golub, L., Hudson, H.~S.,
  Reeves, K.,  \& Warshall, A. 1999, \apj, 527, L121

\bibitem[\protect\citeauthoryear{{Warren} \& {Doschek}}{{Warren} \&
  {Doschek}}{2005}]{warren2005a}
{Warren}, H.~P.,  \& {Doschek}, G.~A. 2005, \apjl, 618, L157

\bibitem[\protect\citeauthoryear{{Winebarger} \& {Warren}}{{Winebarger} \&
  {Warren}}{2004}]{winebarger2004}
{Winebarger}, A.~R.,  \& {Warren}, H.~P. 2004, \apjl, 610, L129

\end{thebibliography}
\end{document}